\documentclass[showpacs,preprintnumbers,amsmath,amssymb,aps,prd,
               nofootinbib,eqsecnum]{revtex4}
\def\be{\begin{equation}}
\def\ee{\end{equation}}
\def\bea{\begin{eqnarray}}
\def\eea{\end{eqnarray}}

\def\R{{\cal{R}}}
\def\S{{\cal{S}}}

\def\dqa{{\delta q_\alpha}}

\def\dq{{\delta q}}
\def\sh{{\sigma_{\rm{s}}}}
\def\Va{{V_\alpha}}
\def\Sab{{{\cal{S}}_{\alpha\beta}}}
\def\Sabo{{\cal{S}}_{\alpha\beta}^{(0)}}

\def\Vab{{V_{\alpha\beta}}}
\def\Vabo{{V_{\alpha\beta}^{(0)}}}
\def\dVab{\dot{V}_{\alpha\beta}}
\def\vp{\varphi}
\def\dvp{{\delta\varphi}}
\def\ri{{\rho_{I}}}
\def\ru{{\rho_{U}}}
\def\Q{{\cal{Q}}}
\def\wt{\widetilde}
\def\dra{{\delta\rho_{\alpha}}}

\def\p{\partial}
\newcommand\eq[1]{Eq.~(\ref{#1})}
\newcommand\eqs[1]{Eqs.~(\ref{#1})}
\begin{document}
\preprint{}
\title{Adiabatic and entropy perturbations with interacting
  fluids and fields} 
\author{Karim A.~Malik$^1$ and David Wands$^2$}
\affiliation{ 
$^1$Physics Department, University of Lancaster, Lancaster LA1 4YB, 
United Kingdom \\
$^2$Institute of Cosmology and Gravitation, University of Portsmouth,
Portsmouth~PO1~2EG, United Kingdom}
\date{\today}
\begin{abstract}
  We develop a gauge-invariant formalism for the study of density
  perturbations in a Friedmann-Robertson-Walker universe with multiple
  interacting fluids and/or scalar fields. We show how $N$ scalar
  fields may be described by $N$ kinetic fluids (with maximally stiff
  equation of state) interacting with a non-dynamical potential (with
  vacuum equation of state). We split generic perturbations into
  adiabatic and entropic parts, and give the coupled first-order
  evolution equations on all scales, including energy and momentum
  exchange. We identify the non-adiabatic effects on large scales,
  and define adiabatic initial conditions in the presence of multiple
  fluids and fields.
\end{abstract}

\pacs{98.80.Cq \hfill JCAP02 (2005) 007, astro-ph/0411703 v2}
\maketitle

\section{Introduction}

The large-scale structure of the cosmos provides the key to
understanding the early evolution of our Universe. In particular
primordial density perturbations may be produced from quantum
fluctuations during an inflationary era.
Most gauge-invariant studies of linear perturbations about homogeneous
and isotropic cosmological background solution follow from the seminal
article of Bardeen \cite{Bardeen}. In particular the reviews of Kodama
and Sasaki \cite{KS} and Mukhanov, Feldman and Brandenberger
\cite{MFB} have become widely used reference works.  An alternative
``covariant'' approach has been advocated by Ellis and collaborators
\cite{Ellis} and developed by Challinor and Lasenby \cite{Challinor}.
These works have provided gauge-invariant descriptions of the
evolution of the total density, pressure and velocity perturbations in
various cosmological models.

Recently there has been increasing emphasis on the evolution of
perturbations in multi-component systems where it is useful to
identify gauge-invariant adiabatic and entropy modes. This is partly
driven by more detailed models of the multi-component systems in the
early universe, such as reheating at the end of inflation
\cite{modreheat}, or the effect of late-decaying scalar fields
\cite{curvaton}. But it is also needed to make more detailed
comparison with observational data in the late-time universe
containing radiation, matter and dark energy (see, e.g., \cite{GordonHu}).

On large scales, where the perturbed universe can be treated as
locally homogeneous and isotropic, we can identify a conserved
perturbation with any locally conserved quantity \cite{conserved}. In
particular there is a conserved perturbation associated with any
perfect fluid whose energy density is locally conserved \cite{WMLL}.
This can be generalised to adiabatic perturbations of a
multi-component cosmology and where the conserved perturbation is the
gauge-invariant curvature perturbation on uniform total density
hypersurfaces, $\zeta$, introduced by Bardeen and collaborators
\cite{BST,Bardeen88}.

Conversely, 
$\zeta$ is not in general conserved
on large scales when one considers non-adiabatic
perturbations of a multi-component system.  Nonetheless, at any
instant, it is possible to decompose arbitrary perturbations of the
multi-component cosmology into an adiabatic mode, with an associated
conserved perturbation on large scales, and entropy (or isocurvature)
modes which evolve independently of the adiabatic mode on large
scales. One consequence of the effect of non-adiabatic perturbations
upon the evolution of the total curvature perturbation, $\zeta$, can
be the generation of correlations between the primordial curvature
perturbation and any surviving isocurvature perturbations at the time
of last-scattering of the cosmic microwave background \cite{Langlois}.

In this paper we will develop the gauge-invariant theory of
cosmological perturbations to deal with multiple interacting fluids
and fields. We develop the multi-component formalism of Kodama and
Sasaki \cite{KS} (see also \cite{hama}). We subtly redefine the
relative entropy perturbations to ensure gauge-invariance when
including energy transfer between fluids. The resulting simplified
evolution equations in the large-scale limit were presented in
Ref.~\cite{mwu}. We also relate this fluid description to previous
work identifying adiabatic and entropy modes in multiple scalar field
systems \cite{chris} (see also
Refs.~\cite{nibbelink,rigopoulos,Hwang:fields,NR,Finelli}). Hwang has
also investigated the evolution of perturbations with multiple fluids
\cite{Hwang:fluids}. See Refs.~\cite{nambu} for an alternative
long-wavelength approximation.

The paper is organised as follows.
In Section \ref{gov_sec} we introduce the governing equations with
multiple interacting fluids in the homogeneous background and for
first-order inhomogeneous perturbations on all scales.  In Section
\ref{gi_perts} we identify gauge-invariant perturbations describing
the total density, velocity and shear perturbations and the relative
entropy and velocity perturbations between different components,
allowing for energy and momentum transfer between the fluids. We give
the coupled first-order evolution equations and in particular give the
simplified set of evolution equations valid in the large-scale limit.
Including interactions enables us to treat fluids and scalar fields on
the same footing and we show in Section \ref{scalarfieldsection} how
$N$ scalar fields can be described as $N$ ``kinetic fluids''
interacting with one ``potential fluid'' describing the vacuum energy.
We conclude in the final section with a short discussion.

In Appendix \ref{gauge_sec} we present the gauge transformation
properties of the metric and matter variables including energy and
momentum transfer and give some identities relating different
variables that prove useful in simplifying our equations. In Appendix
\ref{uniformcurvaturesection} we give the governing equations on all
scales in the uniform curvature gauge which may be better suited to
numerical solutions.

Throughout this work we assume a flat Friedmann-Robertson-Walker (FRW)
background spacetime.  Greek indices, $\mu,\nu,\lambda$, run 
from $0,\ldots3$, while lower case Latin indices, $i,j,k$, run from
$1,\ldots3$. Greek indices from the beginning of the alphabet,
$\alpha,\beta,\gamma$ will be used to denote different fluids, and
upper case Latin indices, $I,J,K$, denote different scalar fields.
We only consider linear scalar perturbations in this article and leave
vector and tensor perturbations to a future publication
\cite{MWreview}.
%

\section{Governing equations}
\label{gov_sec}

In this section we give the governing equations for a system of
multiple interacting fluids.

The covariant Einstein equations are given by
\be
\label{Einstein}
G_{\mu\nu}=8\pi G \; T_{\mu\nu} \,,
\ee
where $G_{\mu\nu}$ is the Einstein tensor, $T_{\mu\nu}$ is the
total energy-momentum tensor, and $G$ is Newton's constant.
Through the Bianchi identities, the field equations (\ref{Einstein})
imply the local conservation the total energy and momentum,
\be
\label{nablaTmunu}
\nabla_\mu T^{\mu\nu}=0\,.
\ee

In the multiple fluid case the total energy-momentum tensor
is the sum of the energy-momentum tensors of the 
individual fluids
\be
T^{\mu\nu}=\sum_\alpha T^{\mu\nu}_{(\alpha)}\,.
\ee
For each fluid we write the local energy-momentum transfer 4-vector as
\be
\label{Qvector}
\nabla_\mu T^{\mu\nu}_{(\alpha)}=Q^\nu_{(\alpha)}\,,
\ee
where energy-momentum is locally conserved, $Q^\nu_{(\alpha)}=0$, only
for non-interacting fluids. 
Equations~(\ref{nablaTmunu}) and~(\ref{Qvector})
imply the constraint
\be
\label{Qconstraint}
\sum_\alpha Q^\nu_{(\alpha)}=0 \,.
\ee

\subsection{Background equations}
\label{background}

The Einstein equations (\ref{Einstein}) give the
Friedmann constraint and evolution equation for 
the background FRW universe
\bea
\label{Friedmann}
H^2 &=& \frac{8\pi G}{3}\rho \,,\\
\label{Hdot}
\dot H &=& -4\pi G \left( \rho+P\right) \,,
\eea
and energy-momentum conservation, Eq.~(\ref{nablaTmunu}), gives the
continuity equation
\be
\label{continuity}
\dot\rho=-3H\left( \rho+P\right)\,,
\ee
where $\rho$ and $P$ are the total energy density and the total
pressure, a dot denotes a derivative with respect to coordinate time,
$t$, the scale factor is $a=a(t)$, and $H\equiv \dot a/a$ is the
Hubble parameter.

The total density and the total pressure are related to the density
and pressure of the component fluids by
\be
\sum_\alpha \rho_{\alpha} =\rho \,, \qquad
\sum_\alpha P_{\alpha} =P \,.
\ee
The continuity equation (\ref{Qvector}) for
each individual fluid in the background is \cite{KS}
\be
\label{cont_alpha}
\dot\rho_{\alpha}
=-3H\left(\rho_{\alpha}+P_{\alpha}\right) +Q_{\alpha}\,,
\ee
where the energy transfer to the $\alpha$-fluid is given by the time
component of the energy-momentum transfer vector
$Q^0_{(\alpha)}=Q_{\alpha}$ in the background.
Equation~(\ref{Qconstraint}) implies that the background energy
transfer obeys the constraint
\be
\label{backconstraint}
\sum_\alpha Q_{\alpha}=0 \,.
\ee

\subsection{Perturbed equations}
\label{pertequations}

We will consider linear scalar perturbations about a spatially-flat FRW
background model, defined by the line element
\be 
\label{ds2}
ds^2=-(1+2\phi)dt^2+2aB_{,i}dt dx^i
+a^2\left[(1-2\psi)\delta_{ij}+2E_{,ij}\right]dx^idx^j \,, 
\ee
where we use the notation of Ref.~\cite{MFB} for the gauge-dependent
curvature perturbation, $\psi$, the lapse function, $\phi$, and scalar
shear, $\sh\equiv a^2\dot E - aB$, and where $\delta_{ij}$ denotes the
flat background metric and $X_{,i}\equiv\p X/\p x^i$.
The transformation of these scalar metric perturbations under
arbitrary gauge-transformations is given in Appendix~\ref{gtrans}.

\subsubsection{Energy and momentum conservation}
\label{energy_and}

The perturbed energy transfer 4-vector, Eq.~(\ref{Qvector}), including
terms up to first order, is written as~\cite{KS,mwu}
\bea
\label{pert_q_vector}
Q_{(\alpha)0} &=& -Q_{\alpha}(1+\phi)-\delta Q_\alpha\,,\nonumber \\
Q_{(\alpha)i}
&=& \left( f_\alpha+Q_\alpha V \right)_{,i} \,,
\eea
and Eq.~(\ref{Qconstraint}) implies that the perturbed energy and
momentum transfer obey the constraints
\be
\label{pertconstraint}
\sum_\alpha \delta Q_{\alpha} = 0 \,,
\quad
\sum_\alpha f_\alpha =0\,.
\ee
The perturbed energy conservation equation for a particular fluid,
including energy transfer, is then obtained by the first-order part of
the time-component of the perturbed continuity equation
(\ref{Qvector}) to give \cite{KS,thesis}
\be 
\label{pertenergyexact}
\dot{\delta\rho}_{\alpha}+3H(\delta\rho_{\alpha}+\delta P_{\alpha})
- 3\left(\rho_{\alpha}+P_{\alpha}\right)\dot\psi
+\frac{\nabla^2}{a^2}(\rho_{\alpha}+P_{\alpha})\left(\Va+\sh\right)
- Q_{\alpha}\phi-\delta Q_{\alpha}=0\,,
\ee
where $\delta\rho_{\alpha}$ and $\delta P_{\alpha}$ are the perturbed
energy density and the perturbed pressure of the $\alpha$-fluid,
respectively, and the comoving spatial Laplacian is denoted by
$\nabla^2\equiv \delta^{ij}\partial^2/\partial x^i\partial x^j$. 
$\Va$ is the covariant
velocity perturbation of the $\alpha$-fluid defined as
\be
\label{defVa}
\Va\equiv a(v_\alpha+B)\,,
\ee
where $v_{\alpha}$ is the scalar velocity potential of the
$\alpha$-fluid.
The momentum conservation equation of the $\alpha$-fluid is
\be
\label{dotVa}
\dot{V}_\alpha
+\left[\frac{Q_\alpha}{\rho_\alpha+P_\alpha}(1+c^2_\alpha)
-3H c^2_\alpha\right]\Va +\phi
+\frac{1}{\rho_\alpha+P_\alpha}\left[
\delta P_\alpha+\frac{2}{3}\frac{\nabla^2}{a^2}\Pi_\alpha
- Q_\alpha V -f_\alpha\right] =0\,,
\ee
where $\Pi_\alpha$ is the perturbed scalar anisotropic stress of the
$\alpha$-fluid and $c^2_{\alpha}\equiv \dot P_{\alpha}/
\dot\rho_{\alpha}$ is the adiabatic sound speed of that fluid.
Note that for stress-free, non-interacting dust we have
$\dot{V}_\alpha+\phi=0$ and hence $\Va=$constant in a synchronous
gauge, whereas the scalar velocity potential redshifts as
$v_\alpha\propto 1/a$.
In Ref.~\cite{mwu} we used $\delta
q_{\alpha}\equiv(\rho_{\alpha}+P_{\alpha})\Va$ as the momentum
perturbation.

The total fluid perturbations are related to the individual fluid
quantities by
\be
\label{sum_delta_rho}
\sum_\alpha \delta\rho_{\alpha} =\delta\rho \,, \qquad
\sum_\alpha \delta P_{\alpha} = \delta P \,, \qquad
\sum_\alpha \Pi_\alpha =\Pi\,,
\ee
and
\be
\label{sum_V}
V=\sum_\gamma\frac{\rho_\gamma+P_\gamma}{\rho+P}V_\gamma\,,
\ee
where $\delta\rho$ is the total density perturbation, $\delta P$ the
total pressure perturbation, $\Pi$ the total anisotropic stress,
and the total covariant velocity perturbation is given by
\be
\label{defV}
V \equiv a(v+B)\,,
\ee
where $v$ is the total scalar velocity potential.

We therefore get an evolution equation for the total density
perturbation from Eq.~(\ref{pertenergyexact}) by summing over all
fluids, using Eq.~(\ref{sum_delta_rho}) and the constraint
(\ref{pertconstraint}),
\be 
\label{pertcontinuity} 
\dot{\delta\rho} + 3H(\delta\rho +
\delta P) -3\left(\rho+P\right)\dot\psi + \frac{\nabla^2}{a^2}
(\rho+P)\left(V+\sh\right) = 0 \,,
\ee
while the total momentum conservation equation is given from
Eq.~(\ref{dotVa}), using Eqs.~(\ref{sum_delta_rho}) and (\ref{sum_V})
and the constraint (\ref{pertconstraint}),
\be
\dot V-3Hc^2_{\rm s} V +\phi
+\frac{1}{\rho+P}\left(\delta
P+\frac{2}{3}\frac{\nabla^2}{a^2}\Pi\right)=0\,,
\ee
where  $c^2_{\rm s}$ is the adiabatic speed of sound, defined as
\be
\label{c2s}
c^2_{\rm s}\equiv\frac{\dot P}{\dot\rho} \,.
\ee
%

\subsubsection{Einstein's field equations}

The $G_0^0$ component of the Einstein equations (\ref{Einstein})
yields the first-order perturbed energy constraint
equation~\cite{KS,MFB,thesis}
\be 
\label{pertFriedmann}
3H\left(\dot\psi+H\phi\right)-\frac{\nabla^2}{a^2}
\left(\psi+H\sh\right) +4 \pi G \delta\rho=0 \,. 
\ee
We get the momentum constraint equation (identically zero in the FRW
background) from the $G^0_i$-component of the Einstein
equations,
\be
\label{momentumconstraint}
\dot \psi + H\phi +4 \pi G (\rho+P)V=0\,.
\ee

{}From the trace-free part of the $G^i_j$ component of 
Einstein's equations we find the shear evolution equation 
\be
\label{shear}
\dot\sh+H\sh-\phi+\psi-8\pi G \Pi=0\,,
\ee
and the trace of the $G^i_j$ component of Einstein's equations gives,
using Eq.~(\ref{shear}), the evolution equation for the curvature
perturbation as
\be
\ddot \psi+3H\dot\psi+H\dot\phi+\left(3H^2+2\dot H\right)\phi
-4\pi G\left( \delta P+\frac{2}{3}\frac{\nabla^2}{a^2}\Pi\right)=0\,.
\ee

\section{Gauge-invariant perturbations and governing equations}
\label{gi_perts}

In this section we define gauge-invariant variables, using the
transformation properties of the metric and matter perturbations given
in Appendix \ref{gtrans}. We then use the results of Section
\ref{pertequations} to get the evolution equations for these
gauge-invariant variables. 

In this and the following sections we will replace spatial Laplacians
by the wavenumbers of their respective eigenmodes according to
$\nabla^2 \to -k^2$.  

\subsection{Definitions}
\label{defvar}

\subsubsection{Curvature perturbations}

A gauge-invariant definition of the curvature perturbation for each
fluid is given by \cite{WMLL}
\be
\label{zetaalpha}
\zeta_{\alpha} \equiv -\psi-H\frac{\delta\rho_\alpha}{\dot\rho_\alpha}
\,.  
\ee 
This describes the dimensionless density perturbation on
uniform-curvature hypersurfaces, or, equivalently at first-order, the
curvature perturbation on uniform $\alpha$-fluid density hypersurfaces
\cite{thesis}.
The total curvature perturbation on uniform density hypersurfaces
\cite{BST,Bardeen88,WMLL} 
\be
\label{defzeta}
\zeta\equiv -\psi-H\frac{\delta\rho}{\dot\rho} \,,
\ee
is then a weighted sum of the individual fluid perturbations
\be
\label{zetatot}
\zeta= \sum_\alpha
\frac{\dot\rho_\alpha}{\dot\rho}  \zeta_\alpha \,.
\ee

We can also define a gauge-invariant combination of the curvature and
velocity perturbations for each fluid
\be
\label{defRalpha}
\R_\alpha\equiv\psi-H\Va\,.
\ee
This describes the curvature perturbation on hypersurfaces orthogonal
to worldlines comoving with the fluid, and is particularly useful when
studying scalar field perturbations.
The total comoving curvature perturbation is given by \cite{Lukash,Lyth1985}
\be
\label{defR}
\R\equiv\psi-HV\,,
\ee
and is again a weighted sum, using Eq.~(\ref{sum_V}), of the
individual fluid perturbations
\be
\label{sumR}
\R
=\sum_\alpha\frac{\rho_\alpha+P_\alpha}{\rho+P}\R_\alpha\,.
\ee

The two alternative descriptions of the total curvature perturbation
are closely related. Indeed even the coefficients
$\dot\rho_\alpha/\dot\rho$ appearing in \eq{zetatot} and
$(\rho_\alpha+P_\alpha)/(\rho+P)$ in \eq{sumR} are the same for
non-interacting fluids, i.e., when $Q_\alpha=0$.
The constraint equation (\ref{pertFriedmann}) can be rewritten
in terms of these gauge-invariant quantities as
\be
\label{Psiconstraint}
\frac{k^2}{a^2}\Psi=3\dot H \left(\R+\zeta\right)\,,
\ee
where we introduce the gauge-invariant curvature perturbation in the
longitudinal or zero-shear gauge \cite{Bardeen,MFB},
\be
\label{defPsi}
\Psi\equiv\psi+H\sh\,.
\ee
In practise it is convenient to use only one of $\zeta$ or $\R$ and
use $\Psi$ to keep track of $\zeta+\R$, which, as can be seen from
(\ref{Psiconstraint}), typically becomes small on super-Hubble scales,
i.e., $k^2/a^2H^2\ll 1$.

The evolution equation for $\zeta_\alpha$ is given from 
\eq{pertenergyexact} and using \eq{pertFriedmann}
\bea
\label{dotzetaalpha1}
\dot\zeta_{\alpha}&=&
3\frac{H^2}{\dot\rho_\alpha}\left(
\delta P_{\alpha}-c^2_\alpha\delta\rho_{\alpha}
\right)
-\dot H\frac{Q_\alpha}{\dot\rho_\alpha}
\left(\frac{\delta\rho_{\alpha}}{\dot\rho_\alpha}-
\frac{\delta\rho}{\dot\rho}\right)
-\frac{H}{\dot\rho_\alpha}\left(
\delta Q_{\alpha}
-\frac{\dot Q_\alpha}{\dot\rho_\alpha}\delta\rho_{\alpha}\right)
\nonumber \\ 
&&+\frac{1}{3H}\frac{k^2}{a^2}\left[
\Psi-\left(1-\frac{Q_\alpha}{\dot\rho_\alpha}\right)\R_\alpha\right]
\,,
\eea
where the terms in the first line are gauge-invariant combinations,
which we shall deal with in the next subsection on entropy perturbations.
The evolution equation for $\R_\alpha$ is given from the momentum
conservation equation, \eq{dotVa}, and using \eq{pertFriedmann}
\bea
\label{dotRalpha1}
\dot\R_\alpha&=&\frac{\dot H}{H}\left(\R_\alpha-\R\right)
-\frac{\dot\rho_\alpha}{\rho_\alpha+P_\alpha}H c^2_\alpha
\left(\R_\alpha+\zeta_\alpha\right)\nonumber\\
&&+\frac{H}{\rho_\alpha+P\alpha}\left[
\delta P_{\alpha}-c^2_\alpha\delta\rho_{\alpha}
-\frac{2}{3}\frac{k^2}{a^2}\Pi_\alpha
-f_\alpha-Q_\alpha\left(V-V_\alpha\right)\right]\,.
\eea

Finally we note that the curvature perturbations on uniform total density,
uniform shear and comoving hypersurfaces, $\zeta$, $\Psi$ and $\R$,
are related to the density, shear and the velocity perturbations on
uniform curvature hypersurfaces,
\be
\label{uni_curv_var1}
\wt{\delta\rho}=-\frac{\dot\rho}{H}\zeta\,, \qquad
\tilde \sh = \frac{\Psi}{H} \,, \qquad \tilde V = -\frac{\R}{H}\,,
\ee
and similarly for the curvature perturbations on uniform
$\alpha$-density and uniform $\alpha$-velocity hypersurfaces
\be
\label{uni_curv_var2}
\wt{\delta\rho_\alpha}=-\frac{\dot\rho}{H}\zeta_\alpha\,, \qquad
\tilde V_\alpha = -\frac{\R_\alpha}{H}\,,
\ee
where the ``tilde'' denotes perturbations on uniform curvature
hypersurfaces.
We can therefore also think of the curvature perturbations $\zeta$ and
$\zeta_\alpha$ as describing density perturbations or density
contrasts, of the curvature perturbations $\R$ and $\R_\alpha$ as
describing velocities, and of the curvature perturbation $\Psi$ as
describing the shear.

Governing equations for the variables defined in \eqs{uni_curv_var1}
and and (\ref{uni_curv_var2}) in uniform curvature gauge are given in
Appendix \ref{uniformcurvaturesection}.

\subsubsection{Entropy perturbations}

The difference between the density perturbation
(\ref{zetaalpha}) for any two fluids describes a gauge-invariant
relative entropy (or isocurvature) perturbation \cite{WMLL,mwu}
\be
\label{defS}
\S_{\alpha\beta} \equiv 3(\zeta_\alpha-\zeta_\beta)
= -3H\left(
\frac{\delta\rho_\alpha}{\dot\rho_\alpha}
- \frac{\delta\rho_\beta}{\dot\rho_\beta} \right) \, .
\ee
Note that this only coincides with the entropy perturbation defined by
Kodama and Sasaki \cite{KS} in the absence of energy transfer,
$Q_\alpha=Q_\beta=0$.
The factor of $3$ is introduced to coincide with the conventional
definition of entropy perturbations in the baryon-photon ratio:
\be
\frac{\delta(n_{\rm{B}}/n_\gamma)}{n_{\rm{B}}/n_\gamma}
= \S_{{\rm{B}}\gamma} \,,
\ee
where $n_{\rm{B}}$ and $n_\gamma$ are the baryon and the photon number
densities, respectively.

Similarly we define the gauge-invariant 
relative velocity perturbation by \cite{KS}
\footnote{ 
The definition of the relative velocity perturbation in
\cite{KS} is related to our definition in \eq{defVab} by 
\be
\Vab=a\Vab_{\rm{KS}}\,.
\ee
}
\be
\label{defVab}
\Vab \equiv \Va-V_\beta = - \frac{1}{H}\left( \R_\alpha-\R_\beta \right) \,.
\ee

The total pressure perturbation can be split into
an adiabatic and non-adiabatic part
\be
\label{splitting_deltaP}
\delta P
\equiv \delta P_{\rm nad}+c^2_{\rm s}\delta\rho\,,
\ee
where $c^2_{\rm s}$ is the adiabatic speed of sound defined in
\eq{c2s} above.

In the presence of more than one fluid,
the total non-adiabatic pressure perturbation, $\delta P_{\rm nad}$,
may be further split into two parts \cite{KS},
\be
\label{deltaPnad}
\delta P_{\rm nad}\equiv \delta P_{\rm intr}+\delta P_{\rm rel}\,.
\ee
The first part is due to the intrinsic entropy perturbation of
each fluid
\be
\label{deltaPintr}
\delta P_{\rm intr}=\sum_\alpha \delta P_{\rm{intr},\alpha} \,,
\ee
where the intrinsic non-adiabatic pressure perturbation of each
fluid is given by
\be 
\label{deltaPintralpha} 
\delta P_{\rm{intr},\alpha}
\equiv \delta P_{\alpha} - c^2_{\alpha}\delta\rho_{\alpha} \,.
\ee
%
%
For any fluid with a definite equation of state, 
$P_\alpha=P_\alpha(\rho_\alpha)$, the intrinsic non-adiabatic pressure
perturbation must vanish, $\delta P_{\rm{intr},\alpha}=0$.
The total adiabatic sound speed (\ref{c2s}) is the weighted sum of the
adiabatic sound speeds of the individual fluids,
\be
c^2_{\rm{s}} =
\sum_\alpha \frac{\dot\rho_\alpha}{\dot\rho} c^2_\alpha \,.
\ee
The second part of the non-adiabatic pressure perturbation
(\ref{deltaPnad}) is due to the {\em relative entropy perturbation}
$\cal{S}_{\alpha\beta}$ between different fluids, defined in
Eq.~(\ref{defS}),
\be
\label{deltaPrel}
\delta P_{\rm rel} =
-\frac{1}{6H\dot\rho}
\sum_{\alpha,\beta}\dot\rho_\alpha\dot\rho_\beta
\left(c^2_\alpha-c^2_\beta\right)\S_{\alpha\beta} \,.
\ee            

Analogous to the non-adiabatic pressure perturbation for each fluid
(\ref{deltaPintralpha}), we can identify an intrinsic non-adiabatic
part of the energy transfer perturbation \cite{mwu} that appears in
the perturbed continuity equation for each fluid
(\ref{pertenergyexact})
\be
\label{deltaQintralpha}
\delta Q_{{\rm intr},\alpha} \equiv \delta Q_\alpha -
\frac{\dot{Q}_\alpha}{\dot\rho_\alpha} \delta\rho_\alpha \,.
\ee
This is automatically zero if the local energy transfer $Q_\alpha$
is a function of the local density $\rho_\alpha$ so that $\delta
Q_\alpha = (\dot{Q}_\alpha/\dot\rho_\alpha)\delta\rho_\alpha$, just
as the intrinsic non-adiabatic pressure perturbation
(\ref{deltaPintralpha}) vanishes when $\delta P_\alpha =
(\dot{P}_\alpha/\dot\rho_\alpha)\delta\rho_\alpha$.
Note however, that from the definition, \eq{deltaQintralpha}, the
intrinsic non-adiabatic energy transfer perturbation, $\delta Q_{{\rm
intr},\alpha}$, can be non-zero in the case where the background
energy transfer $Q_\alpha\neq 0$, even if $\delta Q_\alpha=0$.
We can also identify a relative non-adiabatic energy transfer \cite{mwu}
that appears in Eq.~(\ref{pertenergyexact}) 
whenever $Q_\alpha\neq0$
\be
\label{deltaQrelalpha}
\delta Q_{{\rm rel},\alpha} \equiv
Q_\alpha\frac{\dot H}{H} \left(
\frac{\delta\rho_\alpha}{\dot\rho_\alpha}
-\frac{\delta\rho}{\dot\rho}
\right)
= - \frac{Q_\alpha}{6H\rho} \sum_\beta \dot\rho_\beta \S_{\alpha\beta} \,,
\ee
due to the presence of relative entropy perturbations. Only the
intrinsic parts of the non-adiabatic pressure and energy-transfer
perturbations will appear in the evolution equation for the relative
entropy perturbation, \eq{dotSab} below.

With the above definitions the evolution equation for the curvature
perturbation $\zeta_\alpha$, \eq{dotzetaalpha1}, reduces to
\bea
\label{dotzetaalpha2}
\dot\zeta_{\alpha}&=&
3\frac{H^2}{\dot\rho_\alpha}\delta P_{{\rm intr},\alpha}
-\frac{H}{\dot\rho_\alpha}\left(
\delta Q_{{\rm intr},\alpha}+\delta Q_{{\rm rel},\alpha}\right)
+\frac{1}{3H}\frac{k^2}{a^2}\left[
\Psi-\left(1-\frac{Q_\alpha}{\dot\rho_\alpha}\right)\R_\alpha\right]
\,.
\eea
Thus we see that $\dot\zeta_\alpha\simeq0$ in the absence of
non-adiabatic pressure and energy transfer perturbations, on the large
scales where we can neglect gradient terms.

There is no momentum transfer in the background FRW universe
so the momentum transfer $f_\alpha$ that appears in the momentum
conservation equation for each fluid (\ref{dotVa}) is automatically
gauge-invariant.
Nevertheless we can identify a relative momentum transfer in
\eq{dotVa}, defined as
\be
\label{frelalpha1}
f_{{\rm rel},\alpha} \equiv Q_\alpha\left(V-V_\alpha\right) \,,
\ee
which can be rewritten in terms of the relative velocity perturbation,
defined in \eq{defVab}, using \eq{iden_V}, as
\be
\label{frelalpha2}
f_{{\rm rel},\alpha}
=-Q_\alpha 
\sum_\gamma\frac{\rho_\gamma+P_\gamma}{\rho+P} V_{\alpha\gamma}\,.
\ee
The evolution equation for $\R_\alpha$, \eq{dotRalpha1} then
simplifies to
\bea
\label{dotRalpha2}
\dot\R_\alpha=\frac{\dot H}{H}\left(\R_\alpha-\R\right)
-\frac{\dot\rho_\alpha}{\rho_\alpha+P\alpha}H c^2_\alpha
\left(\R_\alpha+\zeta_\alpha\right)
+\frac{H}{\rho_\alpha+P\alpha}\left[
\delta P_{{\rm intr},\alpha}
-\frac{2}{3}\frac{k^2}{a^2}\Pi_\alpha
-f_\alpha-f_{{\rm rel},\alpha}\right]\,.
\eea

We end up with the following set of gauge-invariant
dynamical variables: the curvature perturbation on uniform density
hypersurfaces, $\zeta$, the curvature perturbation on uniform shear
hypersurfaces $\Psi$, the relative entropy perturbation $\Sab$, and
the relative velocity perturbation $\Vab$.

\subsection{Evolution equations}

In this sub-section we re-express the gauge dependent evolution
equations given in Section \ref{pertequations} in terms of the
gauge-invariant quantities defined above in Section \ref{defvar}. In
deriving the evolution equations, we make extensive use of the
relation between the relative entropy perturbation and the
$\alpha$-fluid and total density perturbations, Eq.~(\ref{iden_rho}),
and the relation between the relative velocity perturbation and the
$\alpha$-fluid and total velocity perturbations, Eq.~(\ref{iden_V}).

The first-order evolution equation for the total curvature
perturbation, $\zeta$, defined in Eq.~(\ref{defzeta}), derives from
the evolution equation of the total density perturbation,
Eq.~(\ref{pertcontinuity}), and is given by
\be
\label{dottotzeta}
\dot\zeta
=-\frac{H}{\rho+P}\delta P_{\rm{nad}}
+\frac{1}{3H}\frac{k^2}{a^2}\left(\Psi-\zeta \right) 
+\frac{1}{9H\dot H}\frac{k^4}{a^4}\Psi\,,
\ee
where $\delta P_{\rm{nad}}$, as defined in Eq.~(\ref{deltaPnad})
depends on the intrinsic entropy perturbations, $\delta P_{{\rm
intr},\alpha}$ defined in Eq.~(\ref{deltaPintralpha}), and the relative
entropy perturbations, $\S_{\alpha\beta}$ defined in
Eq.~(\ref{defS}).

The evolution equation for the curvature perturbation in the
zero-shear gauge, $\Psi$, defined in \eq{defPsi}, follows from
Eqs.~(\ref{momentumconstraint}) and (\ref{shear})
\be
\label{dotPsi}
\dot\Psi+\left(H-\frac{\dot H}{H}\right)\Psi
+\frac{1}{3H}\frac{k^2}{a^2}\Psi
-\frac{\dot H}{H}\zeta=-8\pi GH\Pi\,.
\ee
Note that energy-momentum transfer between individual fluids does not
directly enter into the evolution equations for the total curvature
perturbations $\zeta$ and $\Psi$. However it does considerably
complicate the evolution equations for the relative entropy and
relative velocity perturbations, as can be seen below.

The evolution of the relative entropy perturbation, $\Sab$
defined in Eq.~(\ref{defS}), using \eq{pertenergyexact},
is given by
\bea
\label{dotSab}
\dot\S_{\alpha\beta}
&=&3H\left(\frac{3H\delta P_{\rm{intr},\alpha} 
- \delta Q_{\rm{intr},\alpha}}{\dot\rho_\alpha}
-\frac{3H\delta P_{\rm{intr},\beta} 
- \delta Q_{\rm{intr},\beta}}{\dot\rho_\beta}\right)\nonumber\\
&&+\frac{\dot H}{2H}\left[
\left(\frac{Q_\alpha}{\dot\rho_\alpha}
+\frac{Q_\beta}{\dot\rho_\beta}\right)\Sab
+\left(\frac{Q_\alpha}{\dot\rho_\alpha}
-\frac{Q_\beta}{\dot\rho_\beta}\right)
\sum_\gamma \frac{\dot\rho_\gamma}{\dot\rho} \left(
\S_{\alpha\gamma} + \S_{\beta\gamma} \right)\right]\nonumber\\
&&+\frac{k^2}{a^2}\left[
\left(1-\frac{Q_\alpha}{2\dot\rho_\alpha}
-\frac{Q_\beta}{2\dot\rho_\beta}\right)\Vab
-\left(\frac{Q_\alpha}{\dot\rho_\alpha}
-\frac{Q_\beta}{\dot\rho_\beta}\right)
\left(
\frac{1}{H}\zeta
+\frac{1}{2}
\sum_\gamma\frac{\rho_\gamma+P_\gamma}{\rho+P}
\left(V_{\alpha\gamma}+V_{\beta\gamma}\right)
\right)
\right] \nonumber\\
&&+\frac{k^4}{a^4}\frac{1}{3H\dot H}
\left(\frac{Q_\alpha}{\dot\rho_\alpha}
-\frac{Q_\beta}{\dot\rho_\beta}\right)\Psi
\,.
\eea
Only the intrinsic non-adiabatic perturbations $\delta
P_{\rm{intr},\alpha}$ and $\delta Q_{\rm{intr},\alpha}$ appear
explicitly in the evolution equation (\ref{dotSab}), while the
relative part of the non-adiabatic energy transfer perturbations
(\ref{deltaQrelalpha}) have been expressed in terms of $\Sab$.

Finally, using the evolution equation for the velocity perturbation of the
$\alpha$-fluid, Eq.~(\ref{dotVa}), we can derive an evolution equation
for the gauge-invariant relative velocity perturbation, $\Vab$ defined
in \eq{defVab},
\bea
\label{dotVab}
\dVab&+&\frac{1}{2}\left[
\frac{Q_\alpha}{\rho_\alpha+P_\alpha}
+\frac{Q_\beta}{\rho_\beta+P_\beta}
\right]\Vab 
+\frac{1}{2}\left[\frac{Q_\alpha c^2_\alpha}{\rho_\alpha+P_\alpha}
+\frac{Q_\beta c^2_\beta}{\rho_\beta+P_\beta}
-3H\left(c^2_\alpha+c^2_\beta\right)\right]
\left(\Vab -\frac{1}{3H}\S_{\alpha\beta}\right)\nonumber \\
&&+\frac{1}{2}\left[
\frac{Q_\alpha}{\rho_\alpha+P_\alpha}(1+c^2_\alpha)
-\frac{Q_\beta}{\rho_\beta+P_\beta}(1+c^2_\beta)
-3H\left(c^2_\alpha-c^2_\beta\right)\right]
\sum_\gamma\frac{\rho_\gamma+P_\gamma}{\rho+P}
\left(V_{\alpha\gamma}+V_{\beta\gamma}\right)\nonumber \\
&&-\left[
\frac{Q_\alpha c^2_\alpha}{\rho_\alpha+P_\alpha}
-\frac{Q_\beta c^2_\beta}{\rho_\beta+P_\beta}
-3H\left(c^2_\alpha-c^2_\beta\right)\right]
\left(
\frac{1}{3H\dot H}\frac{k^2}{a^2}\Psi
+\frac{1}{6H}\sum_\gamma\frac{\dot\rho_\gamma}{\dot\rho}
\left(\S_{\alpha\gamma}+\S_{\beta\gamma}\right)
\right)\nonumber \\
&&-\frac{2}{3}\frac{k^2}{a^2}\left(\frac{\Pi_\alpha}{\rho_\alpha+P_\alpha}
-\frac{\Pi_\beta}{\rho_\beta+P_\beta}\right)
-\left(\frac{f_\alpha}{\rho_\alpha+P_\alpha}
-\frac{f_\beta}{\rho_\beta+P_\beta}\right)
+\left(\frac{\delta P_{\rm{intr},_\alpha}}{\rho_\alpha+P_\alpha}
-\frac{\delta P_{\rm{intr},_\beta}}{\rho_\beta+P_\beta}\right)
=0\,,
\eea
where we have used \eq{frelalpha2} for the relative momentum
transfer perturbation.

We therefore see that the evolution equations for $\zeta$, $\Psi$,
$\Sab$ and $\Vab$, \eqs{dottotzeta}-(\ref{dotVab}), form a system of
coupled, first order, ordinary differential equations. Although we
have defined $n(n-1)/2$ relative entropy perturbations, $\Sab$, for
$n$ fluids, there are of course only $n-1$ independent entropy
perturbations, and similarly $n-1$ independent relative velocity
perturbations. Thus we have $2n$ coupled equations relating $2n$
variables describing the density and velocity perturbations in $n$
fluids.

The system is not closed until we specify the intrinsic non-adiabatic
pressure perturbation $\delta P_{\rm{intr},\alpha}$, the intrinsic
non-adiabatic energy transfer perturbation $\delta
Q_{\rm{intr},\alpha}$, the anisotropic stresses $\Pi_\alpha$ and the
perturbed momentum transfer $f_\alpha$ for each fluid. In the
background we have to prescribe the equation of state,
$P_\alpha(\rho_\alpha)$, which determines the adiabatic sound speed
$c^2_\alpha$, and the energy transfer $Q_\alpha$. All these quantities
depend upon the physical model for the fluids. For perfect fluids
$\delta P_{\rm{intr},\alpha}$, $\delta Q_{\rm{intr},\alpha}$,
$\Pi_\alpha$, and $f_\alpha$ are all zero. In Section
\ref{scalarfieldsection} we will calculate these quantities
\emph{explicitly} for the particular example of $N$ scalar fields.

\subsection{Large-scale limit}
\label{largescalesection}

Throughout much of the early history of our Universe, scales of
astrophysical interest are far larger than the Hubble scale, $H^{-1}$,
which defines the cosmological expansion time. For instance, the
Hubble scale at the epoch of primordial nucleosynthesis corresponds to
a comoving scale of around $10$~pc in the present Universe. Thus it is
often a very good approximation to work in a large scale limit when
studying the origin and evolution in the early universe of the large
scale structure of our observable Universe.
In this section we give the governing equations in the large scale
limit, i.e.~in the limit $k\to 0$. Some of the results in this section
have already been presented in \cite{mwu}.

The form of the evolution equations for the perturbations suggests
that the time-dependent amplitude of any perturbation,
$\delta X(t)e^{i{\bf k.x}}$, can be expanded as a Taylor series in
terms of the comoving wavenumber, $k$, as
\be
\label{kexpansion}
\delta X(t) \equiv \sum^\infty_{~~~n=0,1,\ldots} k^{2n} \delta X^{(n)}(t)\,.
\ee
Equations~(\ref{dottotzeta})-(\ref{dotVab}) then give coupled
first-order evolution equations for $X_n(t)$ driven by $X_{n-1}(t)$
for $n\geq1$ [and driven by $X_{n-2}$ for $n\geq2$ in \eqs{dottotzeta}
and~(\ref{dotSab})].

The long-wavelength limit as $k\to0$ is given by the time-dependence
of the $n=0$ solutions to the homogeneous set of first-order equations:
\bea
\label{dotzetatot_lsl} 
\dot\zeta^{(0)} &=& -\frac{H}{\rho+P}\delta P_{\rm nad}^{(0)}
 \,,\\
\label{dotPsi_lsl}
\dot\Psi^{(0)} &=& -\left(H-\frac{\dot H}{H}\right)\Psi^{(0)}
 +\frac{\dot H}{H}\zeta^{(0)} - 8\pi GH\Pi^{(0)}
 \,,\\
\label{S_evol_lsl}
\dot \S_{\alpha\beta}^{(0)} &=& 3H \left(
\frac{3H\delta P_{\rm{intr},\alpha}^{(0)} - \delta
Q_{\rm{intr},\alpha}^{(0)}}{\dot\rho_\alpha}
-\frac{3H\delta P_{\rm{intr},\beta}^{(0)} - \delta
Q_{\rm{intr},\beta}^{(0)}}{\dot\rho_\beta}\right)
+\sum_\gamma \frac{\dot\rho_\gamma}{2\rho} 
\left(
\frac{Q_\alpha}{\dot\rho_\alpha} \S_{\alpha\gamma}^{(0)} 
-\frac{Q_\beta}{\dot\rho_\beta} \S_{\beta\gamma}^{(0)}
\right)
 \,,\\
\label{dotVablarge}
\dot{V}_{\alpha\beta}^{(0)} &=& - \frac{1}{2}\left[
\frac{Q_\alpha}{\rho_\alpha+P_\alpha}
+\frac{Q_\beta}{\rho_\beta+P_\beta}
\right]\Vabo
-\frac{1}{2}\left[\frac{Q_\alpha c^2_\alpha}{\rho_\alpha+P_\alpha}
+\frac{Q_\beta c^2_\beta}{\rho_\beta+P_\beta}
-3H\left(c^2_\alpha+c^2_\beta\right)\right]
\left(\Vabo -\frac{1}{3H}\S_{\alpha\beta}^{(0)}\right)\nonumber \\
&&
- \frac{1}{2}\left[
\frac{Q_\alpha}{\rho_\alpha+P_\alpha}(1+c^2_\alpha)
-\frac{Q_\beta}{\rho_\beta+P_\beta}(1+c^2_\beta)
-3H\left(c^2_\alpha-c^2_\beta\right)\right]
\sum_\gamma\frac{\rho_\gamma+P_\gamma}{\rho+P}
\left(V_{\alpha\gamma}^{(0)}+V_{\beta\gamma}^{(0)}\right)\nonumber \\
&&
+\left[
\frac{Q_\alpha c^2_\alpha}{\rho_\alpha+P_\alpha}
-\frac{Q_\beta c^2_\beta}{\rho_\beta+P_\beta}
-3H\left(c^2_\alpha-c^2_\beta\right)\right]
\frac{1}{6H}\sum_\gamma\frac{\dot\rho_\gamma}{\dot\rho}
\left(\S_{\alpha\gamma}^{(0)}+\S_{\beta\gamma}^{(0)}\right)
\nonumber \\
&&+\left(\frac{f_\alpha^{(0)}}{\rho_\alpha+P_\alpha}
-\frac{f_\beta^{(0)}}{\rho_\beta+P_\beta}\right)
- \left(\frac{\delta P_{\rm{intr},_\alpha}^{(0)}}{\rho_\alpha+P_\alpha}
-\frac{\delta P_{\rm{intr},_\beta}^{(0)}}{\rho_\beta+P_\beta}\right)
 \,.
\eea
This long-wavelength limit exists so long as the inhomogeneous source
terms in the full evolution equations
(\ref{dottotzeta})-(\ref{dotVab}), that is $\delta P_{{\rm
    intr},\alpha}$, $\delta Q_{{\rm intr},\alpha}$, $f_\alpha$ and
$\Pi_\alpha$, all have a well-defined limit as $k\to 0$. 

The perturbation, $\Psi$, vanishes from the other evolution
equations completely in the large scale limit. Since this variable
describes the shear (on uniform curvature hypersurfaces), its
disappearance can be intuitively understood by observing that the
shear is expected to vanish on large scales.
Because the shear and velocity perturbations, $\Psi$ and $\Vab$, only
enter the evolution equations (\ref{dottotzeta}) and (\ref{dotSab}) at
order $k^2$, we can obtain a long-wavelength limit for for $\zeta^{(0)}$ and
$\Sabo$ under the weaker requirement that $\delta P_{{\rm
    intr},\alpha}$, $\delta Q_{{\rm intr},\alpha}$, $k^2f_\alpha$ and
$k^2\Pi_\alpha$ all have a well-defined limit as $k\to0$.
This is the basis of the `separate universes' picture~\cite{WMLL}
commonly used to study the evolution of density perturbations on
sufficiently large scales where the universe looks locally like an
unperturbed (FRW) cosmology. Specifically it requires that we can
neglect the divergence of the momenta in the zero-shear gauge,
$\nabla^2[(\rho_\alpha+P_\alpha)(V_\alpha+\sh)]$, in the perturbed
continuity equation (\ref{pertenergyexact}) for each fluid.

We see from \eq{dotzetatot_lsl} that in the large scale limit the
evolution for the curvature perturbation $\zeta$ is only sourced by
the non-adiabatic pressure perturbation, as defined in \eq{deltaPnad},
including contributions from the intrinsic non-adiabatic pressure
perturbations of each fluid and the relative entropy perturbations
between fluids.
For $\delta P_{\rm nad}=0$ we recover the famous result that
$\zeta=$constant for adiabatic perturbations on large scales
\cite{BST,Bardeen88,GBW,WMLL,conserved,non-pert}.
The long-wavelength solution for adiabatic density perturbations is
\bea
\label{adiabaticsol}
\zeta^{(0)} &=& C \,,\nonumber\\
\Psi^{(0)} &=& \frac{H}{a} \left[ D + \int a \left( \frac{\dot{H}}{H^2}C -
    8\pi G\Pi^{(0)} \right) dt \right] \,,\nonumber\\
\Sabo&=& 0\,.
\eea
where $C$ and $D$ are constants of integration.
This adiabatic solution exists on large scales even in the
presence of energy transfer between fluids, $Q_\alpha\neq0$, so long as
the intrinsic non-adiabatic energy transfer, $\delta Q_{{\rm
    intr},\alpha}$ defined in \eq{deltaQintralpha}, is zero for each
individual fluid.  

The evolution of the relative entropy perturbations, $\Sab$, is
independent of the curvature perturbations $\zeta$ and $\Psi$ in the
large-scale limit, Eq.~(\ref{S_evol_lsl}).
The actual evolution is dependent upon the intrinsic non-adiabatic
pressure and energy transfer of each fluid. However for perfect fluids
all the source terms vanish we have the simple result $\Sab=$constant
in the large-scale limit.

The evolution of the relative velocity perturbations $\Vab$,
\eq{dotVablarge} is driven by relative entropy perturbations, $\Sab$,
as well as the intrinsic non-adiabatic pressure perturbations and
momentum transfer, even on large scales. Only for pressureless,
non-interacting fluids do we find $\Vab=$constant on large scales.

\section{Scalar fields}
\label{scalarfieldsection}

In this section we show how scalar fields can be included in our
interacting fluid formalism. 
Multiple scalar fields provide a test case in which
quantities like the energy transfer and the relative momentum transfer
are simple enough to be computed from first principles, and can then
be applied to identify adiabatic and entropy modes in the fluid formalism.
It also enables us to later ``mix'' fluids and scalar fields in a consistent
way, which is necessary in many cosmological models, including
quintessence models of dark energy, decaying scalar fields, such as
the curvaton, or inhomogeneous reheating at the end of inflation.
In particular we will identify the adiabatic mode and relative
entropy perturbations in the fluid description and compare these with
previous analyses of adiabatic and entropy modes in the scalar field
perturbations~\cite{chris}.

We will split the total energy-momentum of $N$ scalar fields into $N$
maximally stiff fluids interacting with a potential energy. Our
splitting is slightly different from that of Hwang and Noh
\cite{Hwang:fluids} who also relate the scalar field perturbations to
fluid quantities. They work in terms of the total density and pressure
perturbation for each field, including kinetic and potential
perturbations for each field. There is thus an intrinsic non-adiabatic
perturbation for each field. By contrast we separate the kinetic and
potential perturbations so that each has a fixed equation of state and
thus no intrinsic non-adiabatic pressure perturbation and we can
always relate non-adiabatic perturbations to relative entropy
perturbations between different component parts.

For simplicity we consider here only minimally coupled scalar fields
in the sense that they interact only via their combined
self-interaction potential. Although we allow for the presence of
fluids other than the scalar fields, we assume that the scalar fields
do not exchange energy-momentum with any other fluids.

\subsection{Background}

\subsubsection{Standard treatment}

The energy density and the pressure of $N$ minimally coupled scalar
fields, labelled by the subscript ``$I$'', are at zeroth order
\bea
\rho_\vp&=&\frac{1}{2}\sum_I\dot\vp_I^2+U\,, \\
P_\vp&=&\frac{1}{2}\sum_I\dot\vp_I^2-U\,,
\eea
where $U=U(\vp_I)$ is the potential of the scalar fields and upper
case Latin indices, $I,J,K$, run from $1,\ldots,N$.

The adiabatic sound speed of all the scalar fields is given from
\eq{c2s} by
\be
\label{c2s_field}
c^2_{\rm{s}\vp}=1+\frac{2}{3H}\frac{\dot U}{\sum_I\dot\vp_I^2}
 \,,
\ee
where $\dot U \equiv \sum_I U_{,\vp_I}\dot\vp_I$.
%

\subsubsection{Splitting into kinetic and potential parts}
\label{backgroundsplit}

We can split the total energy density and pressure of $N$ scalar
fields into $N$ ``kinetic fluids'' with energy density and pressure,
respectively,
\be
\label{back_ri}
\ri \equiv\frac{1}{2}\dot\vp_I^2 \,, \qquad
P_I\equiv\frac{1}{2}\dot\vp_I^2 \,,
\ee
and a single ``potential fluid'' or vacuum energy,
\be
\label{back_ru}
\ru\equiv U\,, \qquad
P_U\equiv -U\,.
\ee
We now have 
\be
\rho_\vp=\ru+\sum_I\ri\,, \qquad P_\vp=P_U+\sum_I P_I\,.
\ee
The kinetic and potential fluids have a barotropic equation of state
and adiabatic sound speeds
\be
c^2_{{\rm s}I}=1\,, \qquad  c^2_{{\rm s}U}=-1\,.
\ee

The Klein-Gordon equation for each field,
\be
\ddot\vp_I+3H\dot\vp_I+U_{,\vp_I}=0\,,
\ee
gives a continuity equation for each kinetic energy (\ref{back_ru}) of
the form given in Eq.~(\ref{cont_alpha}), where the energy transfer
to the kinetic fluids from the potential is given by
\be
\label{defbackQI}
Q_I=-\dot\vp_I U_{,\vp_I}\,.
\ee
where $U_{,\vp_I}\equiv\partial U/\partial \vp_I$.
The constraint (\ref{Qconstraint}) then implies that 
the energy transfer to the potential is
\be
 Q_U=\sum_I \dot\vp_I U_{,\vp_I}\,.
\ee

\subsection{Perturbations}

We now extend the fluid formalism to the perturbed scalar field
case.

\subsubsection{Standard treatment}

The total energy density and the pressure perturbation for $N$ scalar fields
are
\bea
\delta\rho_\vp&=& \sum_I\left[
\dot\vp_I \dot\dvp_I-\dot\vp_I^2\phi+U_{,\vp_I}\dvp_I\right] \,, \\
\delta P_\vp&=& \sum_I\left[
\dot\vp_I \dot\dvp_I-\dot\vp_I^2\phi-U_{,\vp_I}\dvp_I\right]\,, 
\eea
having contributions both from the individual fields and the potential,
whereas the total velocity perturbation, given by
\be
\label{deftotalV}
V_\vp = -\frac{1}{\sum_K\dot\vp_K^2}\sum_I{\dot\vp_I}{\dvp_I} \,,
\ee
is independent of the potential. 
Scalar fields cannot support anisotropic stress, and we have
\be
\Pi_\vp = 0\,.
\ee

The total non-adiabatic pressure perturbation for $N$ scalar fields is
then readily calculated from \eq{splitting_deltaP} to be
\be
\label{deltaPnadvarphi}
\delta P_{\rm{nad}\vp}
=\frac{2}{3H\sum_K\dot\vp_K^2}\sum_I\dot\vp_I\left[
\ddot\vp_I\delta U-\dot U\left(\dot\dvp_I-\dot\vp_I\phi\right)\right]\,,
\ee
where
\be
\delta U\equiv \sum_I U_{,\vp_I}\dvp_I \,.
\ee

The total entropy perturbation for $N$ scalar fields was denoted $\cal S$ in
Ref.~\cite{chris} and is related to the above definition by
\be
\delta P_{\rm{nad}\vp}=\frac{\dot P_\vp}{H}{\cal S} \,.
\ee
The total entropy perturbation reduces in the single field case to 
\be
\delta P_{\rm{nad}\vp}
=-\frac{2U_{,\vp}\ddot\vp}{3H}\left[
 \frac{\dot\dvp-\dot\vp\phi}{\ddot\vp}
 - \frac{\dvp}{\dot\vp}
\right]\,,
\ee
which is related to the intrinsic entropy perturbation $\Gamma$ defined in
Ref.~\cite{Bartoloetal} by 
\be
\delta P_{\rm{nad}\vp}
=\rho_\vp(1-c^2_{\rm{s}\vp})\Gamma \,,
\ee
and to the intrinsic entropy perturbation $\Gamma_{\rm{q}}$ defined in
Ref.~\cite{Doranetal} by $\delta P_{\rm{nad}\vp}=P_\vp
\Gamma_{\rm{q}}$.

\subsubsection{Splitting into kinetic and potential parts}
\label{perturbationsplit}

Splitting the energy and pressure perturbation, as for the background
fields, into $N$ ``kinetic fluid'' quantities and a single
``potential fluid'' we get
\bea
\label{def_dvpI}
\delta\rho_I&=&\dot\vp_I \dot\dvp_I-\dot\vp_I^2\phi \,, \\
\delta P_I&=&\dot\vp_I \dot\dvp_I-\dot\vp_I^2\phi \,, \\
\Pi_I &=& 0\,,
\eea
for each kinetic fluid, and
\be
\label{def_dru}
\delta\rho_U=\delta U=\sum_I U_{,\vp_I}\dvp_I \,, \qquad
\delta P_U=-\sum_I U_{,\vp_I}\dvp_I \,, \qquad
\Pi_U=0\,,
\ee
for the potential energy. 
We thus have $N$ gauge-invariant curvature perturbations, $\zeta_I$
defined according to \eq{zetaalpha}, for each kinetic fluid 
plus one gauge-invariant curvature
perturbation, $\zeta_U$, for the potential energy.

The covariant velocity perturbation for each kinetic fluid is
\be
\label{def_VI}
V_I = -\frac{\dvp_I}{\dot\vp_I} \,.
\ee
Substituting \eq{def_VI} into the definition of the total velocity,
\eq{sum_V}, we get the standard result, \eq{deftotalV}, for the total velocity
perturbation due to the scalar fields.
Note that the potential energy has no momentum perturbation and hence
its covariant velocity perturbation, $V_U$, is undefined.

The comoving curvature perturbation, defined in Eq.~(\ref{defRalpha}),
for each field is
\be
\label{defRI}
\R_I=\psi+H\frac{\dvp_I}{\dot\vp_I} \,,
\ee
and hence the total comoving curvature perturbation (i.e.~relative to the
average fluid velocity) is given by
\be
\label{Rvp}
\R_\vp = \frac{1}{\sum_K\dot\vp_K^2}\sum_I\dot\vp_I^2\R_I \,.
\ee
The Sasaki-Mukhanov variable \cite{Sasaki1986,Mukhanov}, or the field
fluctuation on uniform curvature hypersurfaces, is simply
$\Q_I=(\dot\vp_I/H)\R_I$.

Note that the field perturbations determine not only the velocity
perturbations (\ref{def_VI}) but also the potential energy perturbation
(\ref{def_dru}). Thus $\delta\rho_U$ and $V_\vp$ are not independent
variables. 
In terms of the gauge-invariant comoving curvature perturbations
$\R_I$ we then have, using \eq{iden_R},
\be
\label{zeta_U2}
\zeta_U=-\frac{1}{\dot U}\sum_K U_{,\vp_K}\dot\vp_K\R_K\,,
\ee
which we can rewrite in terms of gauge-invariant total scalar field comoving
curvature perturbation (\ref{Rvp}) and velocity perturbations
(\ref{deftotalV}) and (\ref{def_VI})
\be
\label{zetaU}
\zeta_U = - \R_\vp 
+ H \sum_I \frac{U_{,\vp_I}\dot\vp_I}{\dot{U}} \left( V_I -V_\vp \right) 
 \,.
\ee
For a single scalar field we have simply $\zeta_U=-\R_\vp$.

Although we have decomposed the total energy-momentum of the $N$
scalar fields into $N$ kinetic fluids interacting with one potential
``fluid'', the potential has no 
momentum perturbation 
and its density perturbation can be written in terms of the kinetic
fluid velocities. Hence we only need the $N$ density perturbations and
$N$ velocity perturbations to describe the perturbed energy-momentum
of $N$ scalar fields.

We get the perturbed energy transfer to each field by substituting
\eq{def_dvpI} for the fluid density into the perturbed energy
conservation equation (\ref{pertenergyexact}), and using the perturbed
Klein-Gordon equation for each field,
\be
\label{KG}
\ddot\dvp_I+3H\dot\dvp_I
+\frac{k^2}{a^2}\left(\dvp_I-\dot\vp_I \sh\right)
+\sum_J U_{,\vp_I\vp_J}\dvp_J
-\dot\vp_I\left(\dot\phi+3\dot\psi\right)
+2 U_{,\vp_I}\phi =0 \,,
\ee
which gives
\be
\label{deltaQI}
\delta Q_I=
-U_{,\vp_I}\left(\dot\dvp_I-\dot\vp_I\phi\right)
-\dot\vp_I\sum_J U_{,\vp_I\vp_J}\dvp_J \,.
\ee
The perturbed energy transfer constraint, (\ref{pertconstraint}),
then gives the energy transfer perturbation to the potential
\be
\delta Q_U=\dot{\delta U}-\dot U \phi \,,
\ee
where
\be
\dot{\delta U}=
\sum_{I,J} \Big[\dot\vp_I U_{,\vp_I\vp_J}\dvp_J \Big]
+\sum_I U_{,\vp_I}\dot\dvp_I \,.
\ee

Finally we identify the momentum transfer to each component.
Substituting the expressions for the $I$-fluid quantities, $V_I$,
$\delta P_I$, $\rho_I+P_I$ and $c^2_I$ of Section
\ref{perturbationsplit} into the momentum conservation equation
(\ref{dotVa}) and, noting that $\Pi_I=0$, we obtain the momentum
transfer perturbation to the $I$-fluid
\be
\label{f_I}
f_I=U_{,\vp_I}\left(\dvp_I+\dot\vp_I V\right)\,,
\ee
in terms of the field variables
or, using \eq{frelalpha1}, 
\be
\label{f_I_relV}
f_I = Q_I\left(V_I-V\right) = -f_{\rm{rel},I} \,.
\ee
Hence the total momentum transfer for each field,
$f_I+f_{\rm{rel},I}$, that appears, for instance, in
Eq.~(\ref{dotRalpha2}), is zero.

As noted earlier the potential fluid has vanishing momentum, and
from the momentum conservation equation (\ref{dotVa}) for the
potential we require
\begin{equation}
f_U = Q_U (V - V_\varphi) \,,
\end{equation}
which is zero if there are no fluids present other than the scalar
fields and $V=V_\varphi$. 
One can verify that Eq.~(\ref{pertconstraint}) for the total momentum
transfer is satisfied with $f_U+\sum_If_I=0$.

\subsection{Relative perturbations}
\label{splitvariables}

Substituting the expression (\ref{def_dvpI}) for the density
perturbation of each 
``kinetic fluid'', and the background density
(\ref{back_ri}), into the definition for the relative density
perturbation, Eq.~(\ref{defS}), we get
\be
\label{S_IJ}
\S_{IJ}=-3H\left[
\frac{\dot\dvp_I}{\ddot\vp_I}-\frac{\dot\dvp_J}{\ddot\vp_J}
-\phi\left(\frac{\dot\vp_I}{\ddot\vp_I}-\frac{\dot\vp_J}{\ddot\vp_J}\right)
\right]\,,
\ee
which is, despite appearances, gauge-invariant as can be seen from the
gauge transformation properties of the variables, Eqs.~(\ref{transphi}) and
(\ref{transrho}). 

The relative density perturbation between each field
and the potential is given by
\be
\label{S_IU}
\S_{IU}=-3H\left[
\frac{\dot\dvp_I-\dot\vp_I\phi}{\ddot\vp_I}-\frac{\delta U}{\dot U}
\right]\,.
\ee
However, remembering that the density perturbation for the potential
can be written in terms of the field velocity perturbations, as in
\eq{zetaU}, we can also write the relative entropy perturbation,
Eq.~(\ref{defS}), between any fluid and the potential as
\be
\label{SalphaU}
\S_{\alpha U}
= 3 (\zeta_\alpha + \R_\vp) -3H \sum_I \left(
\frac{U_{,\vp_I}\dot\vp_I}{\dot{U}} \right) V_{I\varphi} \,.
\ee
In the presence of additional fluids the total comoving curvature
perturbation (\ref{sumR}) 
is given by
\be
\R = \R_\vp - H \sum_\alpha \frac{\rho_\alpha+P_\alpha}{\rho+P}
 V_{\alpha\vp} \,,
\ee
and if we then use \eq{Psiconstraint} to eliminate $\R$ in favour of $\zeta$
and $\Psi$, we finally obtain
\be
\label{SaU}
\S_{\alpha U}=\left( \frac{\dot\rho}{\dot\rho-\dot{U}}\right)
\left[\frac{1}{\dot H}\frac{k^2}{a^2}\Psi
+\sum_{\gamma\neq U} \frac{\dot\rho_\gamma}{\dot\rho}\S_{\alpha\gamma}
-\frac{3H}{\dot U}\sum_{K,\gamma}U_{,\vp_K}\dot\vp_K
\frac{\rho_\gamma +P_\gamma}{\rho+P}V_{K\gamma}\right]\,,
\ee
where we also used \eq{iden_zeta}.

The definition (\ref{S_IJ}) of the relative entropy perturbation,
$\S_{IJ}$, between two fields in terms of their relative density
perturbations, $\delta\rho_I$, differs from the definition used in
Refs.~\cite{chris,nibbelink,rigopoulos}, where the relative entropy
perturbation was defined in terms of the relative {\em field}
perturbations, $\dvp_I$.
In our fluid formalism the relative field perturbation is proportional to
the relative velocity perturbation, defined in Eq.~(\ref{defVab}),
which in the scalar field case, using Eq.~(\ref{def_VI}), is given by
\be
\label{V_IJ}
V_{IJ}=-\left(
\frac{\dvp_I}{\dot\vp_I}-\frac{\dvp_J}{\dot\vp_J}
\right)\,.
\ee
Indeed we can express the relative velocity perturbation, defined
in \eq{V_IJ}, as the difference between comoving curvature perturbations, 
\be
V_{IJ}=-\frac{1}{H}\left(\R_I-\R_J\right)\,,
\ee
which is analogous to the definition of the relative entropy
perturbation $\Sab$ expressed as the difference of curvature perturbations on
uniform density hypersurfaces in \eq{defS}.
We shall see in a moment that when we evaluate the non-adiabatic
energy transfer in terms of the relative perturbations between fields,
the relative velocity perturbation (\ref{V_IJ}) will indeed appear as a
non-adiabatic perturbation.

Note that, as $V_U$ is not defined for the potential energy, the
relative velocity perturbation, $V_{U\alpha}$, between any fluid and
the potential energy is not defined. This would be a problem if
$V_{U\alpha}$ appeared in the evolution equations
(\ref{dottotzeta}--\ref{dotVab}) for our other fluid perturbations.
However, $V_{U\alpha}$ only appears in summations over all fluids,
$\gamma$, where it is multiplied by $\rho_\gamma+P_\gamma$, which
vanishes for the potential energy.
%

\subsection{Non-adiabatic perturbations}
\label{splitvariables2}

Each kinetic fluid and the potential fluid have definite equations of
state ($\delta P_I=\delta\rho_I$, $\delta P_U=-\delta\rho_U$). Hence,
from (\ref{deltaPintralpha}), there is no intrinsic non-adiabatic
pressure perturbation
\be
\label{deltaPintr_IU}
\delta P_{\rm{intr},I}=0\,, \qquad
\delta P_{\rm{intr},U}=0\,,
\ee
And since the kinetic fluids all have the same adiabatic sound speed
($c_{{\rm s}I}^2=1$) there is no relative non-adiabatic pressure
perturbation (\ref{deltaPrel}) due to entropy perturbations between
the different kinetic fluids, $\S_{IJ}$. 
The total entropy perturbation, \eq{deltaPnadvarphi}, is thus due solely to
the relative entropy perturbation between the fields and the potential
(which has $c_{{\rm s}U}^2=-1$):
\be
\label{deltaPnadSIU}
\delta P_{\rm{nad},\vp}
=\frac{2\dot U}{9H^2 \sum_K\dot\vp_K^2}\sum_I \dot\vp_I\ddot\vp_I\S_{IU}\,.
\ee
Note, however, that this entropy perturbation can always be
re-expressed in terms of a sum over the other entropy perturbations and
relative velocity perturbations, \eq{SalphaU}.

The intrinsic non-adiabatic energy transfer perturbation, defined in
\eq{deltaQintralpha}, for the $I$-fluids follows from \eqs{back_ri},
(\ref{defbackQI}), (\ref{def_dvpI}), and (\ref{deltaQI}),
and is given by
\be
\delta Q_{\rm{intr},I}
=\dot\vp_I\left[
\frac{\dot U_{,\vp_I}}{\ddot\vp_I}\left(\dot\dvp_I-\dot\vp_I\phi\right)
-\delta U_{,\vp_I}\right]\,,
\ee
which can be rewritten as
\be
\delta Q_{\rm{intr},I}
=\sum_J \dot\vp_I U_{,\vp_I\vp_J} \left[ -\dvp_J +
  \frac{\dot\vp_J}{\ddot\vp_I} \left(\dot\dvp_I-\dot\vp_I\phi\right)
\right] \,.
\ee
This is a relative entropy perturbation between the proper time
derivative of the $I$-th field and the value of the $J$-th field. 
It is convenient to expand this as
\be 
\delta Q_{\rm{intr},I} 
=\sum_J \dot\vp_I \dot\vp_J U_{,\vp_I\vp_J}
\left[ \frac{\delta U}{\dot{U}} -\frac{\dvp_J}{\dot\vp_J} +
\frac{\left(\dot\dvp_I-\dot\vp_I\phi\right)}{\ddot\vp_I} -
\frac{\delta U}{\dot{U}} \right] \,, 
\ee 
because, using \eq{V_IJ}, we can write
\be
\frac{\delta U}{\dot{U}} -\frac{\dvp_J}{\dot\vp_J} 
= \sum_K
\frac{U_{,\vp_K}\dot\vp_K}{\dot{U}} \left[ \frac{\dvp_K}{\dot\vp_K} -
\frac{\dvp_J}{\dot\vp_J} \right] 
= \sum_K
\frac{U_{,\vp_K}\dot\vp_K}{\dot{U}} V_{JK} \,.  
\ee 
So finally, using \eq{S_IU}, we can express the non-adiabatic
intrinsic energy transfer in terms of the relative density
perturbations $\S_{IU}$ and velocity perturbations $V_{IK}$
\be
\label{deltaQintrSIU_I} 
\delta Q_{\rm{intr},I} =\sum_J \dot\vp_I \dot\vp_J U_{,\vp_I\vp_J}
\left[ \sum_K \frac{U_{,\vp_K}\dot\vp_K}{\dot{U}} V_{JK} -
\frac{1}{3H} \S_{IU} \right] \,.  
\ee
For the U-fluid the intrinsic non-adiabatic energy transfer
perturbation is given by
\be
\label{deltaQintrU}
\delta Q_{\rm{intr},U}
=\dot{\delta U}-\dot U\phi-\frac{\ddot U}{\dot U}\delta U \,,
\ee
where we have used \eqs{deltaQintralpha}, (\ref{back_ru}),
(\ref{defbackQI}), and (\ref{def_dru}), which yields in terms of
relative entropy perturbations
\be
\label{deltaQintrSIU_U}
\delta Q_{\rm{intr},U}
=- \frac{1}{3H} \sum_I U_{,\vp_I}\ddot\vp_I\S_{IU}
+\sum_{I,J}U_{,\vp_I\vp_J}\dot\vp_I\dot\vp_J
\frac{\sum_K U_{,\vp_K}\dot\vp_K}{\dot{U}}V_{KJ}\,.
\ee

Thus we see that a non-zero relative velocity perturbation, $V_{JK}$,
between scalar fields leads to a non-adiabatic energy
transfer in Eqs.~(\ref{deltaQintrSIU_I}) and (\ref{deltaQintrSIU_U}).
To recover the adiabatic solution (\ref{adiabaticsol}) on
long-wavelengths with $\S_{IU}=0$ we see, from
Eq.~(\ref{S_evol_lsl}), that we require in addition that $V_{JK}\to0$ for
scalar fields with arbitrary interaction potential $U(\vp_I)$.  Thus
we can identify the relative velocity perturbation
$V_{JK}$ between scalar fields as a non-adiabatic perturbation
\cite{chris}, in contrast to the case of perfect fluids where we
require only that $k^2\Vab\to0$ as $k\to0$ to obtain an adiabatic
solution with $\Sab=0$ on large scales.

We can finally rewrite the momentum transfer (\ref{f_I_relV}) solely in
terms of a sum over the relative velocity perturbations $V_{I\alpha}$,
according to \eq{frelalpha2},
\be
\label{f_I3}
f_I
 =
 - \dot\vp_I U_{,\vp_I}
 \sum_\alpha\frac{\rho_\alpha+P_\alpha}{\rho+P}V_{I\alpha}
 \,.
\ee

\section{Discussion and conclusions}

In this paper we have extended the analysis in the classic paper by
Kodama and Sasaki \cite{KS} to deal with the coupled evolution of
curvature and isocurvature perturbations in a multi-component
cosmology with interacting fluids and scalar fields. In doing so we
have clarified the nature of adiabatic initial conditions, and
identified non-adiabatic effects.

It is well-known that the total curvature perturbation $\zeta$ is
conserved on large scales when the total non-adiabatic pressure
perturbation (\ref{splitting_deltaP}) vanishes. In a multi-component
system the total non-adiabatic pressure perturbation can be split,
according to Eq.~(\ref{deltaPnad}), into the sum of the intrinsic
non-adiabatic pressure perturbations of individual components and the
relative pressure perturbation due to the relative density
perturbations, $\S_{\alpha\beta}$, between different components. The
intrinsic non-adiabatic pressure perturbations depend on the internal
degrees of freedom of the fluid, but must vanish for fluids with a
definite equation of state $P_\alpha(\rho_\alpha)$.

The relative density perturbation between different components,
$\S_{\alpha\beta}$ defined in (\ref{defS}), generalises the
perturbation in the photon-baryon ratio in a standard hot big bang
cosmology to arbitrary interacting fluids. Hence we refer to
$\S_{\alpha\beta}$ as a relative entropy (or isocurvature) perturbation.

We have given, for the first time, the evolution equations on all
scales for the gauge-invariant entropy perturbations
$\S_{\alpha\beta}$ in a multi-component system, which is driven by the
intrinsic non-adiabatic pressure perturbations and perturbed energy
transfer.  Analogously to the pressure perturbations we split the
perturbed energy transfer into the gauge-invariant intrinsic
non-adiabatic energy transfer and the relative energy transfer, due to
the relative entropy perturbations, $\S_{\alpha\beta}$. In certain
models, such as spontaneous decay of non-relativistic particles
\cite{mwu}, one may also be able to give the intrinsic non-adiabatic
energy transfer in terms of the relative entropy perturbations,
enabling one to obtain a closed set of evolution equations.

The overall curvature perturbation and relative entropy perturbations
are also coupled to the velocity perturbations, but these decouple in
the large scale limit (in the absence of intrinsic non-adiabatic
pressure and intrinsic non-adiabatic energy transfer).
Hence adiabatic density perturbations are characterised by a single
amplitude, $\zeta$, which remains constant on large scales.
A sufficient condition for adiabatic perturbations on
large scales is 
\be
\delta P_{{\rm intr},\alpha}=0\,, \qquad 
\delta Q_{{\rm intr},\alpha}=0\,, \qquad  
{\rm{and}} \qquad\S_{\alpha\beta}=0\,,
\ee 
for all components $\alpha$ and $\beta$. 
Adiabatic perturbations then stay adiabatic on large scales.

We have also shown how to describe scalar fields and their perturbations in
this notation. $N$ interacting scalar fields can be described by $N$
kinetic ``fluids'' with a stiff equation of state, $P_I=\rho_I$,
interacting with one potential ``fluid'' with vacuum equation of
state, $P_U=-\rho_U$. The potential fluid velocity is undefined,
because its momentum perturbation $(\rho_U+P_U)V_U$ is zero.  Because
the components have fixed equations of state there are no intrinsic
non-adiabatic pressure perturbations, only the relative non-adiabatic
pressure due to relative entropy perturbations. On the other hand
there is an intrinsic non-adiabatic energy transfer which includes
terms due to the relative velocity perturbation, $V_{IJ}$, even on
large scales. Thus adiabatic perturbations for interacting scalar
fields on large scales require 
\be
S_{IJ}=0\,, \qquad S_{IU}=0\,, \qquad {\rm{and}} \qquad V_{IJ}=0\,.
\ee

Each scalar field perturbation $\delta\vp_I$ determines the velocity
perturbation, $V_I$, and thus adiabatic field perturbations
\cite{chris} must obey
\begin{equation}
V_{IJ} =\frac{\delta\vp_J}{\dot\vp_J}
-\frac{\delta\vp_I}{\dot\vp_I}= 0 \,.
\end{equation}
During inflation scalar field perturbations originate as small-scale
quantum fluctuations and these will not in general respect this adiabatic
condition. On the other hand the existence of a unique attractor in
phase space at late times would drive perturbations towards $V_{IJ}=0$.
In addition the adiabatic condition, $\S_{IJ}=0$, for relative density
perturbations requires that
\begin{equation}
\frac{\delta\dot{\vp}_{I\tau}}{\ddot\vp_I} =
 \frac{\delta\dot{\vp}_{J\tau}}{\ddot\vp_J} \,,
\end{equation}
where $\delta\dot{\vp}_{I\tau}\equiv \dot{\delta\vp}_I-\dot\vp_I\phi$
is the perturbed proper time derivative of the field, and $\S_{IU}=0$
requires
\begin{equation}
\frac{\delta\dot{\vp}_{I\tau}}{\ddot\vp_I} = \frac{\delta U}{\dot{U}} \,.
\end{equation}

All of the above conditions can be enforced by requiring the {\em
  generalised} adiabatic condition for perturbations in {\em any}
4-scalars $x$ and $y$ \cite{WMLL}:
\begin{equation}
\label{gen_condition}
\frac{\delta x}{\dot x} = \frac{\delta y}{\dot y} \,.
\end{equation}
This is a generalisation of the usual adiabatic condition,
$\S_{\alpha\beta}=0$, for fluid density perturbations which requires
\begin{equation}
\frac{\delta\rho_\alpha}{\dot\rho_\alpha} =
\frac{\delta\rho_\beta}{\dot\rho_\beta} \,.
\end{equation}

Because the scalar field perturbation determines both the velocity
perturbation and the potential perturbation, $\S_{IU}$, is not
independent of the other variables. For instance, for a single
scalar field we have, from Eq.~(\ref{SaU}), 
\begin{equation}
\S_{IU} =\left( \frac{\dot\rho}{\dot\rho-\dot{U}}\right)
\left[\frac{1}{\dot H}\frac{k^2}{a^2}\Psi
+\sum_{\gamma\neq U} \left(\frac{\dot\rho_\gamma}{\dot\rho}\S_{I\gamma}
 -3H \frac{\rho_\gamma +P_\gamma}{\rho+P} V_{I\gamma}\right) \right] \,.
\end{equation}
In a cosmology dominated by the scalar field this forces the scalar
field perturbations to become adiabatic, $\S_{IU}\to0$, in the $k\to0$
limit for finite $\Psi$ \cite{chris}. But in the presence of other
fluids, adiabatic field perturbations require the scalar field
velocity to coincide with the total velocity perturbation, and hence
$\sum_\gamma(\rho_\gamma+P_\gamma)V_{I\gamma}=0$.

Our formalism is applicable to a variety of cosmological models
including interacting scalar fields and fluids.
Although we have only considered minimally coupled scalar fields our
formalism can naturally be extended to non-minimally coupled 
fields. Conformally transforming to the Einstein frame introduces
interactions between a scalar field and the matter, of the form
\cite{Damour}
\begin{equation}
Q_{(I)}^\mu \propto T_m \nabla^\mu \vp_I \,,
\end{equation}
where $T_m$ is the trace of the matter energy-momentum tensor. On the
other hand the perturbed decay of an oscillating massive scalar field
into light particles may be described by the perturbed energy transfer
between pressureless matter and radiation \cite{modreheat}.

\acknowledgments

The authors are grateful to David Lyth for useful comments.
KAM is supported by PPARC grant PPA/G/S/2002/00098.
DW is supported by the Royal Society.

\appendix

\section{Gauge transformations and some useful equations}
\label{gauge_sec}

\subsection{Gauge transformations}
\label{gtrans}

The metric tensor, including linear scalar perturbations about a flat
FRW background is given by
\be
g_{\mu\nu}= \left( 
\begin{array}{cc} 
-1-2\phi & a B_{,i} \\ 
a B_{,j} & a^2\left[(1-2\psi)\delta_{ij} + 2E_{,ij} \right]
\end{array} 
\right) \,.
\ee
A first order coordinate transformation 
$\tilde x^\mu=x^\mu+\delta x^\mu$,
where $\delta x^\mu=[\delta t, \delta x_,^{\,i}]$,
induces a change in the metric tensor
\be
\widetilde {g_{\mu\nu}}=g_{\mu\nu}-\pounds_{\delta x^\mu} 
g_{\mu\nu}\,\
\ee
where $\pounds_{\delta x^\mu}$ denotes the Lie-derivative with respect
to $\delta x^\mu$.

The scalar functions $\phi$, $\psi$, $B$ and $E$ change under the
transformation as
\begin{eqnarray}
\label{transphi}
\widetilde\phi&=&\phi-\dot{\delta t} \, ,\\
\label{transpsi}
\widetilde\psi&=&\psi+\frac{\dot a}{a}\delta t \, , \\
\label{transB}
a\widetilde B&=&aB-a^2\dot{\delta x}+\delta t \, ,\\
\label{transE}
\widetilde E&=&E-\delta x \, .
\label{scaltran}
\end{eqnarray}

The total four velocity is subject to the constraint
\be
u_\mu u^\mu=-1 \,,
\ee
and allowing for linear perturbations we find
\bea
u_\mu&=&\left[-(1+\phi),a(v+B)_{,i}\right]\,, \\
u^\mu&=&\left[1-\phi,\frac{1}{a}v_,^{~i}\right]\,.
\eea
A first order coordinate transformation 
induces a change in the four velocity according to
\be
\widetilde u_\mu=u_{\mu}-\pounds_{\delta x^\mu} 
u_{\mu}\,.
\ee
In a similar fashion we can define fluid four velocities for the
individual fluids.
We then find that the velocity potentials transform as
\be
\tilde v=v+a\dot{\delta x}\,, \qquad 
\tilde v_\alpha =v_\alpha+a\dot{\delta x}\,,
\ee
and hence the covariant scalar velocity perturbations, 
defined above in \eqs{defVa} and (\ref{defV}), as
\bea 
\Va&\equiv& a\left(v_\alpha+B\right) \,, \nonumber \\
V&\equiv& a\left(v+B\right) \,, \nonumber
\eea
transform as
\bea
\widetilde\Va &=& \Va+\delta t\,, \\
\widetilde V &=& V+\delta t\,.
\eea

Note, in \cite{mwu} we used momentum perturbations 
$\dq$ and $\dqa$ instead of the velocity perturbations
$V$ and $\Va$ which are related by
\bea 
\dqa&=&(\rho_\alpha+P_\alpha)\Va \,, \\
\dq&=&(\rho+P)V \,.
\eea
The total and the individual momenta change
as
\be
\widetilde{\dq}=\dq+(\rho+P)\delta t\,, \qquad 
\widetilde{\dqa}=\dqa+(\rho_\alpha+P_\alpha)\delta t\,.
\ee
The shear scalar, $\sh\equiv a^2\dot E-aB$, changes as
\be
\tilde\sh=\sh-\delta t\,.
\ee
The density perturbation changes as
\be
\label{transrho}
\widetilde{\delta\rho}=\delta\rho-\dot\rho\delta t\,.
\ee
The perturbed energy transfer four vector of the $\alpha$-fluid,
$Q^\nu_{(\alpha)}$, defined in \eq{pert_q_vector}, transforms as
\be
\widetilde Q^\nu_{(\alpha)}=Q^\nu_{(\alpha)}-\pounds_{\delta x^\mu} 
Q^\nu_{(\alpha)}\,,
\ee
and we find that the perturbed energy transfer to the $\alpha$-fluid
changes as
\be
\label{transQ}
\widetilde{\delta Q_\alpha}=\delta Q_\alpha-\dot Q_\alpha\delta t\,,
\ee
whereas the perturbed momentum transfer is gauge-invariant
\be
\widetilde{f_i}=f_i\,.
\ee

\subsection{Identities}

\subsubsection{Background}

A useful relation between the Hubble parameter and the background
energy density can be derived by combining Eqs.~(\ref{Friedmann}),
(\ref{continuity}), and (\ref{Hdot}) to give
\be
\frac{\dot H}{H}=\frac{\dot \rho}{2\rho}\,.
\ee

\subsubsection{Perturbations}

In order to derive the set of evolution equations for $\Sab$ and
$\Vab$, \eqs{dotSab} and (\ref{dotVab}), we made use of the following
relations between the $\alpha$-fluid and total density perturbation
and the relative entropy perturbation,
\be
\label{iden_rho}
\frac{\delta\rho_\alpha}{\dot\rho_\alpha}
=\frac{\delta\rho}{\dot\rho}
-\frac{1}{3H}\sum_\gamma\frac{\dot\rho_\gamma}{\dot\rho}
\S_{\alpha\gamma}\,,
\ee
and the relation between the relative velocity perturbation and the
$\alpha$-fluid and total velocity perturbation,
\be
\label{iden_V}
\Va=V+\sum_\beta\frac{\rho_\beta+P_\beta}{\rho+P}\Vab\,.
\ee
In terms of the curvature perturbations on uniform density and
comoving hypersurfaces the above expressions give
\bea
\label{iden_zeta}
\zeta_\alpha&=&\zeta+\frac{1}{3}\sum_\gamma\frac{\dot\rho_\gamma}{\dot\rho}
\S_{\alpha\gamma}\,, \\
\label{iden_R}
\R_\alpha&=&\R- H\sum_\beta\frac{\rho_\beta+P_\beta}{\rho+P}\Vab\,.
\eea

We found it useful to ``symmetrise'' some expressions involving
differences in the $\Vab$:
\be
x_\alpha \sum_\gamma\frac{\rho_\gamma+P_\gamma}{\rho+P} V_{\alpha\gamma}
-x_\beta\sum_\gamma \frac{\rho_\gamma+P_\gamma}{\rho+P} V_{\beta\gamma}
\equiv
\frac{1}{2}\left(x_\alpha+ x_\beta\right)\Vab
+\frac{1}{2}\left(x_\alpha- x_\beta\right)
\sum_\gamma\frac{\rho_\gamma+P_\gamma}{\rho+P}\left(V_{\alpha\gamma}+
V_{\beta\gamma}
\right)\,.
\ee
A similar expression holds for $\Sab$, replacing the
$\rho_\gamma+P_\gamma$ and $\Vab$ by $\dot\rho_\gamma$ and $\Sab$.

\section{Evolution equations in the uniform curvature gauge}
\label{uniformcurvaturesection}

Uniform-density or uniform-field gauges, used to define the curvature
perturbations $\zeta_\alpha$ and $\R_I$, can become ill-defined in
some cases even though nothing singular is happening in other
gauges. This occurs whenever the uniform field hypersurfaces become
singular, i.e.~if $\dot\vp_I=0$, as can be seen from \eq{defRI}
\cite{Finelli2,llmw}, or when the uniform density hypersurfaces become
singular, which happens when $\dot\rho_\alpha=0$, as can be seen from
\eq{zetaalpha}.  In the latter context the equations presented below
were used in \cite{gmw}, but only in their large scale limit.
The singular behaviour can be eliminated by simply working with
rescaled perturbation variables which physically corresponds to
working with perturbations in the uniform-curvature gauge
\cite{KS}. 
The density contrast $\Delta_{\rm{g}}$ defined in \cite{KS} and
denoted $\Delta$ in \cite{Doranetal} is related to the density
perturbation on flat slices by
$\Delta_{\rm{g}}=\Delta=\wt{\delta\rho}/\rho$. 

Note that all the variables in this section are defined on uniform
curvature hypersurfaces. To avoid confusion we denote quantities
evaluated in the uniform curvature gauge by a ``tilde''. 
The curvature perturbations $\zeta$, $\zeta_\alpha$, $\R$ and
$\R_\alpha$ are simply related to the density and velocity
perturbations in the uniform-curvature gauge via
Eqs.~(\ref{uni_curv_var1}) and (\ref{uni_curv_var2}) .

In the case of multiple fluids the energy evolution equation of the
$\alpha$-fluid in the uniform curvature gauge is given from
\eq{pertenergyexact} by
\be 
\label{dotdeltarhoalphacur}
\dot{\wt\dra}+\left[
3H\left(1+c^2_\alpha\right)-\frac{\dot Q_\alpha}{\dot\rho_\alpha}
\right]\wt\dra
-\delta Q_{{\rm intr},\alpha}+\delta P_{{\rm intr},\alpha}
- \frac{k^2}{a^2}(\rho_{\alpha}+P_{\alpha})(\wt{\Va}+\wt{\sh})
-Q_{\alpha}\frac{\dot H}{H}\wt{V}=0\,.
\ee
The evolution equation for the velocity 
perturbation of the $\alpha$-fluid, $\Va$, is, using \eq{dotVa}
\be
\label{evolVacur}
\dot{\wt\Va}+\left[\frac{Q_\alpha}{\rho_\alpha+P_\alpha}(1+c^2_\alpha)
-3Hc^2_\alpha\right]\wt\Va
+\frac{\dot H}{H}\wt{V}
+\frac{1}{\rho_\alpha+P_\alpha}\left[
\delta P_{{\rm intr},\alpha}+ c^2_\alpha\wt\dra
-\frac{2}{3}\frac{k^2}{a^2}\Pi_\alpha-Q_\alpha\wt V-f_\alpha\right] =0\,.
\ee
The total velocity and the velocities of the individual species
are related by \eq{sum_V},
whereas the total density perturbation is just the sum
of the density perturbations of the individual fluids, \eq{sum_delta_rho}.
Summing over the evolution equations of the individual fluids,
\eq{dotdeltarhoalphacur}, and using the constraints
\eqs{sum_delta_rho} and (\ref{sum_V}), we find the total energy
evolution equation to be
\be 
\label{dotdeltarhocur} 
\dot{\wt{\delta\rho}} + 3H(1+c^2_{\rm s})\wt{\delta\rho} 
+3H \delta P_{\rm nad} 
-\frac{k^2}{a^2}\left[(\rho+P)(\wt{V}+\wt{\sh})\right] = 0 \,,
\ee
and the total momentum evolution equation from \eq{evolVacur},
\be
\dot{\wt{V}}+\left(\frac{\dot H}{H}-3Hc^2_{\rm s}\right)\wt{V}
+\frac{c^2_{\rm{s}}}{\rho+P}\wt{\delta\rho}
+\frac{1}{\rho+P}\left(
\delta P_{\rm{nad}} -\frac{2}{3}\frac{k^2}{a^2}\Pi\right)=0\,.
\ee
The shear in the uniform curvature gauge, $\tilde\sigma$, is simply
the rescaled metric potential $\Psi$, defined in \eq{defPsi}, as can
be seen from \eq{uni_curv_var1}, so to close the system of equations
we can either use the constraint equation (\ref{Psiconstraint})
together with \eq{uni_curv_var1}, or use the shear evolution equation
in the uniform curvature gauge,
\be
\dot{\wt{\sh}}+H\wt{\sh}-\frac{\dot H}{H}\wt{V}-8\pi G\, \Pi=0\,.
\ee
%



\end{document}